\documentclass[12pt]{article}
\usepackage[left=3cm, right=3cm, top=5cm, bottom=5cm]{geometry}
\usepackage{graphicx} 
\usepackage{amsmath, amssymb, verbatim}
\usepackage{booktabs}
\usepackage{xcolor}
\usepackage{url}
\usepackage{array}
\usepackage{hyperref}
\usepackage{bm}
\usepackage{epstopdf}
\usepackage{multirow}
\usepackage{subfigure}
\usepackage{authblk}
\usepackage{algorithm}
\usepackage{algorithmic}

\newtheorem{teo}{Theorem}

\title{\textbf{Nonparametric Bayesian analysis for the Galton-Watson process }}

\author[1]{Massimo Cannas}
\author[2]{Michele Guindani}
\author[3 *]{Nicola Piras}

\affil[1]{Department of Economic and Business Sciences, University of Cagliari}
\affil[2]{Department of Biostatistics, University of California, Los Angeles}
\affil[3]{Department of Mathematics and Computer Science, University of Cagliari}

\affil[ ]{*Corresponding author: nicola.piras97@unica.it}
\affil[ ]{Contributing authors: massimo.cannas@unica.it; mguindani@g.ucla.edu}
\date{}
\begin{document}
\maketitle

\begin{abstract}
The Galton-Watson process is a model for population growth which assumes that individuals reproduce independently according to the same offspring distribution. Inference usually focuses on the offspring average as it allows to classify the process with respect to extinction. We propose a fully non-parametric approach for Bayesian inference on the GW model using a Dirichlet Process prior. The prior naturally generalizes the Dirichlet conjugate prior distribution, and it allows learning the support of the offspring distribution from the data as well as taking into account possible overdispersion of the data. The performance of the proposed approach is compared with both frequentist and Bayesian procedures via simulation. In particular, we show that the use of a DP prior yields good classification performance with both complete and incomplete data. A real-world data example concerning COVID-19 data from Sardinia illustrates the use of the approach in practice.
\end{abstract}

\section{Introduction}
\label{sec:1}
The Galton-Watson (GW) model is a stochastic model originally developed by Francis Galton and Henry William Watson in 1875 to investigate the extinction of aristocratic surnames in Victorian England \cite{WatsonGalton1875}. It was then used by Lotka to analyze US census data to understand population dynamics, using the zero-inflated geometric distribution to model offspring distribution patterns \cite{lotka}. He calculated a high extinction probability, suggesting that the vast majority of family lines were destined to disappear over time, while population sustainability depended on a relatively small proportion of continuing lineages. The model was then adapted by  R.A. Fisher to analyze several genetic phenomena, like the survival and extinction patterns of mutant genes in large populations, where he used the process to model genetic drift, that is, the random changes in allele frequencies across generations. A concise historical review can be found in the classic monograph of Harris~\cite{harris}. Since then, the model has seen several applications in many fields. For example, in the 1930s, the Hungarian physicist Leo Szilard developed the process to describe nuclear chain reactions; a theoretical framework for modeling the extinction probability for a neutron fission chain which is still employed today \cite{FerreiradeAraujoRoberty2024, Tantillo2024}. Additional applications include modeling the cascades of neural activations in the brain ~\cite{Pausch2020}, information propagation through peer-to-peer systems and online social systems ~\cite{DraiefMassoulie2011, Gleeson2020} and disease spread in epidemiology ~\cite{Spouge2019}. The latter is the domain of application we are also considering in this work. 
 
The GW process tracks the evolution of a population across discrete generations, either by recording the total number of individuals in each generation or, when complete data are available, by tallying counts of individuals by their offspring size. Standard inference schemes for this process typically assume a parametric offspring distribution with a finite, fixed support. Examples include Poisson, binomial, or truncated distributions. However,  such rigid assumptions can miss complex features like multimodality, overdispersion, or heavy tails. To overcome these limitations, we propose using a Dirichlet Process (DP) prior on the unknown offspring distribution. The  flexibility of the DP nonparametric prior allows for an effectively infinite support and lets the observed data determine the number and weights of distinct reproduction probabilities. Moreover, thanks to the conjugacy properties of the DP, one can carry out posterior inference via efficient algorithms, allowing to take into account the probabilistic uncertainty in the estimation process of the characteristics of the process. 

The paper is organized as follows. In section \ref{sec:2}, we review the main characteristics of the GW process as well as some major contributions in the statistical literature for inference on the GW process. In particular, we discuss the prior proposed by Heyde \cite{heyde} for the so-called Poisson-GW, which leads to a simple expression for the posterior of the offspring average, that is, the expected number of offspring per individual, and the non-parametric prior proposed by Mendoza and Guti\'errez-Pe\~na ~\cite{gutpena}, which leads to approximate expressions for the posterior of the offspring average. In Section~\ref{sec:3}, we propose a fully nonparametric prior based on the Dirichlet process. In Section \ref{sec:4}, we compare the performance of our proposal with existing methods via a simulation study. We further presents results from applying our model in a case study using COVID-19 data from the Italian region of Sardinia in Section~\ref{sec:5}. Finally, Section~\ref{sec:6} provides some concluding remarks.

\section{A review of the Galton-Watson process and related inference}
\label{sec:2}

 The GW process assumes that individuals in a population reproduce independently, and that each individual gives rise to a random number of descendants according to an \textit{offspring distribution}, 
\begin{eqnarray*}
 P(\textrm{an individual has $j$ descendants}) = \pi_j, \quad j\in \mathbb{S},  
\end{eqnarray*}
where $\mathbb{S}$ denotes the finite or (countably) infinite support of the offspring distribution. A key quantity is the \textit{offspring average} (or mean reproduction number), that is,  the expected number of individuals generated by each individual,
$$m = \displaystyle\sum_{j\in \mathbb{S}} j\pi_j.$$
For example, if the offspring distribution is Poisson $(\lambda)$, then
$\pi_{j}=e^{-\lambda} \lambda^{j}/{j!}$ and $ m=\sum_{j=0}^{\infty} j \pi_{j}=\lambda$, so  $\lambda$ plays the role of the average number of offspring per individual. 

Let $Z_0$ indicate the original ancestors of a population. Then,  the evolution of the population is characterized by the process $(Z_i)_{i\in \mathbb{N}}$, where $Z_i$ denotes the total number of individuals in the $i$-th generation. In the following, without loss of generality, we assume $Z_0=1$. The GW process is Markovian; hence, the future evolution of the process does not depend on previous generations given that we know the size of the last generation. In particular, the state $0$ is absorbing: if $Z_i = 0$ for some $i$ then $Z_r=0$ for all $r > i$. In case $Z_i=0$ we say that the population is \textit{extinct} at time $i$. Correspondingly,   the \emph{extinction probability} can be defined as the probability that the population  eventually dies out, i.e., as $q=P\left(\exists i: Z_{i}=0\right)$. 
Intuitively, since a larger offspring average $m$ drives faster expected growth, increasing $m$ drives down the value of $q$. The link between the offspring distribution and the extinction probability is precisely described in the following classic result \cite{WatsonGalton1875, Feller1968},
\begin{teo}(Extinction probability)\label{propclassic}
Let $G(s)$ be the probability generating function of the offspring distribution. Then,  extinction probability of the GW process is the smallest root of the equation 
$$G(q) = q. $$ 
Moreover,  $q<1 \Longleftrightarrow m>1$, where $m$ is the average of the offspring distribution.
\end{teo}
The proposition above lets us classify three regimes by comparing the offspring mean
$m$ with the critical threshold \(1\). In particular, we distinguish \emph{subcritical} (\(m<1\)) and \emph{critical} (\(m=1\)) processes, which both die out almost surely (\(q=1\)), from \emph{supercritical} (\(m>1\)) processes, which survive with positive probability (\(q<1\)). In many cases one can solve the fixed-point equation \(G(q)=q\) in closed form. For example, if the offspring distribution is geometric,
\(
P(X=j) = p\,(1-p)^j,\quad j\ge0,
\)
then its mean is
\(
m = {(1-p)}/{p},
\)
and the extinction probability is
\(
q =1\) if $ m \le 1$ and $q=1/m$ if $m > 1$. In contrast, for a Poisson\((\lambda)\) offspring law, 
\(
G(s) = e^{\lambda(s-1)},
\)
the extinction probability in the supercritical case \(\lambda>1\) is the unique root \(q<1\) of
\(
e^{\lambda(q-1)} = q,
\)
which typically must be solved numerically or expressed using the Lambert \(W\)-function. More generally, heavy-tailed or power-law offspring distributions yield generating functions involving polylogarithms, and one finds \(q\) by solving \(G(q)=q\) approximately.

Since the extinction in a Galton-Watson process depends fundamentally on the offspring law, inference typically focuses on the offspring-distribution mean \(m = \sum_{j\in\mathbb{S}} j\,\pi_j\). Other parameters of interest include the probabilities \(\pi_j\) and the support size \(\lvert\mathbb{S}\rvert\). The statistical literature typically distinguishes two observation schemes. Under
incomplete data, one observes only the sequence of generation sizes \(Z_0,Z_1,\dots,Z_n\), where \(n\) is the number of generations. In contrast, complete-data additionally record the counts \(Z_{ij}\) of the number of individuals in generation \(i\) having exactly \(j\) offspring. Therefore, for each generation \(i\),  \(Z_i = \sum_{j\in\mathbb{S}} Z_{ij}\) and the next generation can be obtained as \(Z_{i+1} = \sum_{j\in\mathbb{S}} j\,Z_{ij}\). Correspondingly, the total sample size can be computed as \(N = \sum_{i=0}^n Z_i = \sum_{i=0}^n\sum_{j\in\mathbb{S}} Z_{ij}\).\\

\noindent
{\bf Maximum Likelihood based inference.}
Under mild regularity conditions on the offspring distribution, Harris~\cite{harris} showed that the maximum-likelihood estimator of the  offspring average \( m \), based on complete data up to generation \( n \), is simply the total number of children divided by the total number of parents
$$\qquad \hat{m}=\frac{Z_1+\dots +Z_n}{Z_0+\dots +Z_{n-1}}.$$ This estimator is consistent, provided the process does not become extinct. Interestingly, the same estimator holds under incomplete data, when only the generation totals \( Z_{0},\dots,Z_{n-1} \) are observed. Indeed, letting \(\hat{\pi}_j\) denote the maximum-likelihood estimator of \(\pi_j\), one obtains $\hat{m}(\{Z_i\}) = \sum_{j} j\hat{\pi}_j$, and
  \begin{align*}
   \sum_{j} j\hat{\pi}_j
  =0\cdot\frac{\sum_i Z_{i0}}{\sum Z_i}+\cdots + k\cdot \frac{\sum_i Z_{ik}}{\sum Z_i}
  =\frac{\sum_j j \sum_i Z_{ij}}{Z_0+\cdots Z_{n-1}} =\frac{\sum_i\sum_j jZ_{ij}}{Z_0 + \cdots Z_{n-1}}=\frac{Z_1+\cdots Z_n}{Z_0 + \cdots Z_{n-1}}.
  \end{align*}
  This equality follows from the invariance property of maximum-likelihood estimators and the sufficiency of the observed totals; see Keiding and Lauritzen~\cite{lauritzen}.\\

\noindent
{\bf Bayesian inference.} 
Bayesian inference for the GW process has been considered by several authors. Here, we recall the analyses of Heyde \cite{heyde} and Mendoza and Guti\'errez-Pe\~na ~\cite{gutpena}, which we will later compare to our approach. Heyde~\cite{heyde}, who assumed incomplete data and a power series offspring distribution, showed that the improper prior \(\pi(m)=m^{-1}\) yields a gamma posterior distribution in the special case of Poisson offspring. He further derived an approximation for general power series offspring distributions, such that
\[
P(m > 1) \approx P\left(\chi^2_{2(Z_n - Z_0)} > 2Z_{n-1}\right).
\]

A Bayesian conjugate analysis for the GW process was proposed by Mendoza and Guti\'errez-Pe\~na ~\cite{gutpena}. Assuming complete data and a finite-support offspring distribution with known support size $k+1$, they adopted a conjugate Dirichlet prior distribution for the offspring probability distribution $\pi= (\pi_0,\pi_1,\dots,\pi_k) \sim Dirichlet (\alpha_0,\dots,\alpha_k)$,  leading to the posterior
$$\pi|\{Z_{ij}\} \sim Dirichlet (\beta_0,\dots,\beta_k) \qquad \textrm{where} \quad  \beta _j=\alpha _j+\sum_{i=0}^n Z_{ij}.$$
In other words, each prior parameter $\alpha_j$ is updated adding the total count of individuals  producing exactly $j$ offspring across generations.  Consequently,
each component $\pi_{j}$ follows a beta distribution, both a priori and a posteriori. For example,  with the uniform (flat) prior $\left(\alpha_{0}, \ldots, \alpha_{k}\right)=(1, \ldots, 1)$, the posterior distribution of $\pi_j$ is
\begin{equation}
  \pi_j|\{Z_{ij}\}\sim Beta(\beta_j,\beta-\beta_j)\equiv 
Beta\left(1+\sum_i Z_{ij}, k+\sum_{ij} Z_{ij} -\sum_i Z_{ij} \right). 
\label{eq:01}
\end{equation}
Inference on the mean offspring number $m$
exploits the linear relationship $m=\mathbf{h}^{\prime} \pi$, where $\mathbf{h}=(0,1, \ldots, k)^{\prime}$, i.e. the offspring average is a linear combination of the entries of the vector $\pi$. Hence,  the first two posterior moments of $m$ can be approximated using the moments of $\pi$:
\begin{equation*}
E(m|\{Z_{ij}\})=\textbf{h}'\mu=m(\mu) \quad Var(m|\{Z_{ij}\})=\textbf{h}'\,\Sigma\,\textbf{h}=\sigma^2(\mu)/(\beta+1)
\end{equation*}
with $\mu$ and $\Sigma$ indicating the posterior mean vector and the variance-covariance matrix of $\pi$. Note that \(m\) is a linear combination of beta-distributed random variables, and its exact posterior distribution is generally not available. Therefore, inference for \(m\) typically relies on asymptotic approximations. Specifically, a normal approximation for the posterior distribution is valid under the assumption of non-extinction of the process. However, the accuracy of this approximation in small-sample scenarios should be evaluated through simulations. Additionally, Mendoza and Guti\'errez-Pe\~na ~\cite{gutpena} proposed a beta approximation for the posterior distribution of \(m\), which leads to an uninformative prior for \(m\). We further discuss this prior in the Appendix.

\section{A Dirichlet Process prior on the offspring distribution}
\label{sec:3}

\noindent 
Over the past two decades, the Dirichlet Process (DP), introduced by Ferguson \cite{ferguson}, along with its various extensions, has been extensively used in data analysis due to its flexibility.  A significant feature of the DP is its ability to
approximate any complex distribution as a mixture of potentially infinitely many components, without requiring to fix the number of parameters in advance and with minimal prior information. Under mild conditions, any generative model can be accurately approximated by the DP, provided it shares the same support as the base measure of the process. We refer for details to M\"uller et al. \cite{quintana}, who provide and overview of Bayesian nonparametric methods and applications. Guindani et al. \cite{guindani} proposed a DP-based approach for the analysis of count data from genetic experiments. The GW model produces count data, representing either the total number of individuals in each generation or the number of individuals with a fixed offspring size. Thus, it is natural to employ the DP as a prior for the offspring distribution, since it allows for a fully nonparametric specification of the offspring distribution. In contrast to parametric or finite-support models, the DP allows the support of the offspring distribution to be countably infinite and learned from the data. This avoids arbitrary truncations and enables more robust inference in cases where the true support is unknown or potentially large. Moreover, the discreteness of the DP ensures that the posterior concentrates on a finite subset of offspring values actually observed, making it a natural choice for modeling uncertainty in the offspring law.

We start by assuming complete data. Let $G$ denote the unknown offspring distribution, then we assume that the observed data 
\begin{equation}\label{eq1} 
  \{Z_{ij}\} | G \sim  G. 
\end{equation}
The DP prior allows to model the distribution $G$ flexibly,  by representing it as an infinite sum of weighted point masses \cite{sethuraman1994}. Different characterizations of the DP can be provided. For our purposes, it is convenient to refer to the constructive definition in \cite{ferguson}, whereby $G$ is defined as a random probability measure such that, for any finite measurable partition $(A_0, A_1, \ldots, A_k)$ of the real line, the finite-dimensional vector 
\begin{equation*}
(G(A_0), G(A_1), \ldots, G(A_k)) \sim \text{Dirichlet}(aG_0(A_0), \ldots, aG_0(A_k)),
\end{equation*}
where $a > 0$ is a concentration (or precision) parameter that determines how closely $G$ adheres to a base measure $G_0$, which represents prior beliefs about the offspring distribution. In symbols, $G \sim DP(a,G_0).$ Intuitively, albeit somewhat imprecisely, one could say that the above DP prior is centered around a parametric distribution $G_0$ for the offspring distribution with degree of precision $a$. Indeed, for any measurable set $A$, the mean and variance of $G(A)$ are
\(
\mathbb{E}[G(A)] = G_0(A), \quad \mathrm{Var}(G(A)) = G_0(A)\big(1 - G_0(A))/{(a + 1)},
\)
highlighting the role of the parameter $a$ as a precision parameter.  In practice, $G_0$ is a  prior parameter (a distribution)  representing the researcher's belief about the offspring distribution. The parameter $a$ regulates the degree of confidence in such belief, with larger values associated with increased confidence. For example, a strong belief that the data may come from a Poisson-GW can be accommodated by setting $G_0$ Poisson and a large value of $a$. Instead, if the value of $a$ is small, the prior allows realization to vary widely from any particular base offspring distribution. \\

When we assign the DP prior to the offspring distribution, we implicitly assign a prior on $\pi_j$, $j \in \mathbb{N}$. To elaborate, consider a partition of the positive real line where each $A_j = [j, j+1)$ for $j \in \mathbb{N}$. Then, by definition, the DP prior induces a distribution on the probability vector $\pi = (G(A_0), \ldots, G(A_k))$ for any $k \in \mathbb{N}$. More specifically, $\left(G\left(A_{0}\right), G\left(A_{1}\right), \cdots G\left(A_{k}\right)\right) \sim \operatorname{Dirichlet}\left(a G_{0}\left(A_{0}\right), \ldots, a G_{0}\left(A_{k}\right)\right)$. This allows us to see the Dirichlet prior of Mendoza and Guti\'errez-Pe\~na \cite{gutpena} as a special case of the DP prior specification considered here: a $\pi \sim \text{Dirichlet}(\alpha_0, \ldots, \alpha_k)$ prior is equivalent to $\pi$ derived from $G \sim \text{DP}(a, G_0)$ where $a = \sum_j \alpha_j$ and $G_0(A_j) = \alpha_j / \sum_j \alpha_j$. Conversely, if $\pi$ is distributed as $G \sim \text{DP}(a, G_0)$ where $G_0$ has support contained in $\{0,...,k+1\}\subset \mathbb{N}$, then $\pi$ follows a $\text{Dirichlet}(aG_0(A_0), \ldots, aG_0(A_k))$ distribution. This relationship arises directly from the definition of the Dirichlet Process and provides a natural interpretation of the DP as a nonparametric extension of the Dirichlet distribution.

For example, the flat Dirichlet prior with $(\alpha_0,\cdots, \alpha_k)=(c,\cdots,c)$ is equivalent to a $DP(a,G_0)$ when $a=c(k+1)$ and $G_0$ is the uniform discrete distribution on $\{0,1,\cdots, k\}$. Indeed $aG_0(A_j)=c(k+1)Pr(G_0=j)=c$ so marginally $\pi_j\sim Beta(a G_0(A_j),a(1-G_0(A_j))\equiv Beta(c,ck)$ as in the conjugate parametric analysis. The flat Dirichlet prior is also equivalent to $DP(1,G_0)$ where $G_0\sim U(0, k+1)$. More generally, $DP(a, G_0)$ with $G_0$ discrete can be also expressed as $DP(a,C_0)$ where $C_0$ is a continuous probability measure such that $C_0(A_j)=G_0(A_j)$.\\ 

{\bf Posterior inference.} One of the key advantages of the Dirichlet Process is its conjugacy property: the posterior distribution, after observing data, remains a Dirichlet Process with updated parameters. This property greatly simplifies Bayesian computation, as posterior inference and predictive distributions can be derived in closed form. Indeed, if $G \sim DP(a,G_0)$ be a Dirichlet Process with base measure $G_0$ and precision $a$, the posterior is 
\begin{equation}
\label{eq3}
G|\{Z_{ij}\}\sim DP\left(a+N,\frac{1}{N+a} \sum_{j \in \mathbb{S}}\sum_{i=1}^n \delta_{z_{ij}}+\frac{a}{N+a}G_0\right),
\end{equation}

 with an updated base measure given by a weighted average of the prior guess $G_0$ and the empirical distribution of the observed sample, and an updated concentration parameter $a+N$. Here, $N$ denotes the total number of individuals observed in the sample. As is typical in Bayesian analysis, the influence of the prior is greater when the sample size is small, such as in subcritical processes or when only a few generations are observed. For $A \subset \mathbb{S}$ we have
\begin{equation}
E(G(A) | \{Z_{ij}\})=\frac{N}{N+a}  \sum_{j \in A}\sum_{i=1}^n \frac{1}{N}\delta_{z_{ij}}(A)+\frac{a}{N+a}G_0(A)
\end{equation}
If we take $A=[j,j+1)$ the previous formula can be interpreted saying that, a posteriori, the expected proportion of type $j$ individuals is a weighted average of the observed proportion of type $j$ individuals in the sample and the proportion of individuals of type $j$ according to $G_0$.

 Inference on $m$ can be obtained from posterior summaries such as $E(G|\{Z_{ij}\})$ or $\text{median}(G|\{Z_{ij}\})$. In addition to estimating $m$ and extinction probabilities, the DP prior allows inference on the size of the offspring support. Specifically, the support size can be inferred by inspecting the posterior distribution of the number of distinct values assumed by realizations of the process.  Owing to the discrete nature of the DP, ties are expected in such samples. This is a property that underlies the clustering behavior of the DP and has been widely used in species sampling and text classification, among other applications \cite{quintana}. Antoniak \cite{antoniak} derived the distribution of $K_N$, the number of unique values in a sample of size $N$, and showed that $K_N$ grows asymptotically with the logarithm of $N$. This distribution exhibits a ``rich get richer" dynamic, where observations tend to tie together in large groups as the sample size grows. However, as Miller and Harrison \cite{miller} have shown for data generated from a finite mixture, the posterior distribution of $K_N$ may not always converge to the true value. More recent results in Ascolani et al. \cite{ascolani} indicate that if the parameter $a$ is assigned a suitable prior with bounded support or a proper Gamma prior, then the posterior distribution of $K_N$ can be consistent for the true number of mixture components when analyzing data generated from finite mixtures. Nevertheless, this consistency holds only as the sample size approaches infinity, which in our context is only possible for supercritical processes. In our simulations, we examine the small-sample behavior of inference for both $m$ and the offspring support size (see Section~\ref{sec:4}).

\subsection{Agnostic DP prior}\label{NeutralDP}
It is important to elicit the DP prior to be agnostic, or neutral, in the sense that it does not favor any particular classification of the process. For example, such a prior would assign equal probability to the process being critical or subcritical, one half to each, and thus, a priori, would not favor either extinction or explosion of the process. For the classic Dirichlet prior on $\pi_0, \dots, \pi_k$, Mendoza and Guti\'errez-Pe\~na ~\cite{gutpena} showed that an uninformative prior for $m$ can be obtained by setting 
\begin{equation*}
\alpha_0=1,\alpha_1\approx 0, \cdots, \alpha_{k-1}\approx 0 ,\alpha_k=\frac{log(2)}{log(k)}
\end{equation*}
This agnostic prior is obtained imposing that the induced distribution on $m$ has median equal to one; see the Appendix for details. Given that the Dirichlet prior construction can be seen as a special case of the DP, an equivalent DP prior assumes  $G\sim DP(a, G_0)$ with $a=1+log(2)/log(k)$ and $G_0=[1/a,0\cdots,0,(a-1)/a]$. Another equivalent prior is a $DP(a, G_0)$ with $G_0$ distributed as a non standard beta distribution on the interval $[0, k]$. 

We note that the agnostic priors discussed above are applicable only when $k$ is finite. However, for infinite support, we can still establish an agnostic Dirichlet Process prior by imposing the constraint that $\text{median}(G_0)=1$. Ferguson \cite{ferguson} demonstrated that the median of a sample from $\text{DP}(a, G_0)$ is the median of $G_0$ itself. Therefore, any $\text{DP}(a,G_0)$ where $\text{median}(G_0)=1$ provides an agnostic prior specification. Table~\ref{tab1} summarizes several choices of $G_0$ corresponding to common offspring distributions, along with the parameter values that render the prior uninformative for the mean $m$. For instance, with $G_0$ following a Poisson($\lambda$) distribution, imposing $\text{median}(G_0)=1$ yields the equation $\bigl\lfloor \lambda+\frac{1}{3}-\frac{1}{50\lambda}\bigr\rfloor=1$, which is satisfied when $\lambda \approx 0.6954$.

\begin{table}[ht] \centering
\footnotesize
\noindent 
\resizebox{\textwidth}{!}{
\begin{tabular}{|c|c|}
\hline 
{$G_0$} & {Parameters for agnostic prior}  \\[8pt]
\hline

Discrete on $\{0,\dots,k\}$ & $G_0(0)=\dfrac{1}{a},G_0(1)=\dots = G_0(k-1)=0,G_0(k)=\dfrac{a-1}{a}; \quad a=\frac{log(2)}{log(k)}$  \\[8pt]
\hline 

Poisson($\lambda$) & 
$\lambda\approx 0.6954$ \\[8pt]
\hline 

Geometric($\lambda$) & 
$\lambda\approx 1-\frac{\sqrt{2}}{2}\approx 0.2928$ \\[8pt]
\hline
\end{tabular}
}
\caption{Parameters of $G_0$ giving an agnostic prior for the offspring average $m$.}\label{tab1}
\end{table}

\subsection{DP prior with incomplete data}
\label{algoincomplete}
Let us now consider the case where only the vector of incomplete data $\{Z_{i}\}$ is available. Gonzales et al.~\cite{gonzalez} developed a Bayesian analysis framework for
  GW type processes using a Dirichlet prior under incomplete data. To address this limited information scenario, they proposed a blocked Gibbs sampler that simultaneously estimates both the missing complete data and the offspring average.
   The algorithm exploits the fact that the conditional distribution of $Z_{ij}$ given $Z_i$ is Multinomial. The sampling procedure begins by initializing the offspring distribution $\pi^\ell$ in step $0$ by setting $\pi^{0}\sim Dirichlet(\beta_0^0,\dots,\beta_k^0)$ the algorithm iterates the following steps until convergence: 
\begin{description}
\item[1)] generate $\{Z_{ij}\}^{\ell}|(\{Z_{i}\},\pi^{\ell-1}) \sim Multinomial(\{Z_{i}\},\pi^{\ell-1})$ subject to $\displaystyle \sum_{j=0}^k jZ_{ij}=Z_{i+1} $;\\
\item[2)] generate $ \pi^{\ell}|\{Z_{ij}\}^{\ell}\sim Dirichlet(\beta_0^\ell,\dots,\beta_k^\ell)$.\\
\end{description}
The constraint in Step~1 ensures that the complete data generated in iteration $\ell$ is consistent with the observed row total at generation $\ell+1$. This consistency requirement necessitates implementing an accept-reject sampling procedure for Step~1, where candidates are drawn from the multinomial distribution and accepted only if they satisfy the summation constraint.
 
The algorithm can be adapted to work with our fully nonparametric approach. To incorporate the Dirichlet Process prior for the incomplete data scenario, we need only modify Step~2 of the algorithm. Instead of sampling the parameter vector $\pi^{(\ell)}$ from the Dirichlet distribution, we now sample from the posterior distribution of $G$ given the imputed complete data $\{Z_{ij}\}^{(\ell)}$. The adapted algorithm is then described in \textbf{Algorithm 1}.

\begin{algorithm}
\caption{Sampling Algorithm for GW Process with DP Prior and Incomplete Data}
\begin{algorithmic}[1]
\STATE Initialize $\pi^{(0)} \sim \text{DP}(a, G_0)$
\WHILE{not converged}
\STATE \textbf{Step 1:} Generate $\{Z_{ij}\}^{(\ell)}|(\{Z_{i}\},\pi^{(\ell-1)}) \sim \text{Multinomial}(\{Z_{i}\},\pi^{(\ell-1)})$ subject to 
    $\displaystyle \sum_{j=0}^k jZ_{ij}=Z_{i+1}$ 
\STATE \textbf{Step 2:} Update posterior $\pi^{(\ell)} \sim \text{DP}(a + N, \frac{1}{N+a}\sum_{j=1}^k \sum_{i=1}^{n} \delta_{z_{ij}}+\frac{a}{N+a}G_0)$, \\
where $N= \text{ total number of individuals}$
\ENDWHILE
\RETURN Posterior samples of $\pi$ and $\{Z_{ij}\}$
\end{algorithmic}
\end{algorithm}


Step~1 requires  knowledge of $k$. This can be estimated at each iteration by computing $K^\ell$, the number of unique values in the samples $Z_{ij}$. However, since the $Z_{ij}$ are updated at each iteration, this strategy may lead to poor mixing of the chain. It is often much simpler and computationally efficient to set a large value of $k$, i.e. assume a truncated DP prior. We observe that the implicit accept-reject procedure in step~1 represents a potential computational bottleneck, as generating a consistent row may become extremely time-consuming for certain parameter configurations.  Interestingly, our simulation studies indicate that the DP-adapted version suffers less from this drawback. This improved efficiency likely stems from the more flexible posterior distribution provided by the Dirichlet Process, which can better adapt to the observed data patterns and thus increase the probability of generating acceptable samples that satisfy the required constraints.



\section{Simulation}
\label{sec:4}
\noindent In this section, we present a Monte Carlo simulation study comparing the performance of the Dirichlet Process prior against the Bayesian inferential methods discussed in Section~\ref{sec:2}. Our simulation focuses on evaluating small sample performance of these methods, which is particularly relevant for practical applications. This focus is deliberate since the consistency properties of the compared estimators, while theoretically valid, typically require large datasets to manifest;  specifically, supercritical processes observed across multiple generations, a scenario rarely encountered in applied settings. An additional objective of our simulation is to assess whether the $DP$ prior can recover the true offspring size when the offspring distribution has finite support.\\ 

\noindent
{\bf Data generation. } We generated the data from a GW process starting at $Z_0=1$. We considered several scenarios for the offspring distribution: i) finite support and ii) infinite support. For each scenario we generated 500 samples from subcritical, critical and supercritical processes. In each scenario we assume that the process can be observed for the first 10 generations.\\  

\noindent
{\bf Evaluation Metrics} The performance of the estimation methods is assessed calculating the proportion of correct classifications over simulated samples. We classify the process as (sub)critical or not by comparing estimates of $m$ with the threshold value of one. More specifically, we use the following approach for each method:
\begin{itemize}
\item Maximum likelihood estimator. We compare the maximum likelihood estimate of $m$ with the threshold value of one: if $\hat{m}<1$ the process is considered (sub)critical, otherwise (super)critical.

\item Bayesian analysis with improper prior \cite{heyde}. We approximate $P^*=P(m>1 |  \, data)$, the posterior probability that the average offspring is greater than one. Then we classify the process as (sub)critical if $P^*< 0.5$ and (super)critical otherwise. 

\item Bayesian analysis with non-informative Dirichlet prior ~\cite{gutpena}. We compare  $\hat{m}=\sum_{j=1}^{k} j \hat{\pi_j}$ with the threshold value of one and classify, using the same classification rule as for the maximum likelihood estimator.  
\end{itemize}

For the DP prior approach, we compare $\hat{m}=E(G_0 \mid \text{data})$ with the threshold value of one, classifying the process as (sub)critical if $\hat{m}<1$ and (super)critical otherwise. An additional advantage of the DP prior is that it enables estimation of the offspring support size. This estimate is derived as the modal size of samples obtained from the DP posterior distribution.\\


\noindent
{\bf Prior parameters.} For our simulation study, we specified the prior parameters for each Bayesian approach as follows. For the Bayesian conjugate analysis with the  Dirichlet prior of \cite{gutpena}, we employed the non-informative parameters as described earlier. A Bayesian analysis using Heyde's improper prior requires no choice of parameters. For the Bayesian analysis with the $\text{DP}(a,G_0)$ prior, we investigated both low and high values of the concentration parameter $a$. Low values allow to ``learn" the offspring from the data. We also investigated what happens with high values of $a$ reflecting an high confidence on the central measure $G_0$. In all cases using the DP prior, we selected $G_0$ as a Poisson distribution with parameter calibrated to yield an agnostic prior where $P(m>1)=0.5$, as described in Section~\ref{NeutralDP}. We always set $G_0$ as Poisson even when the prior has a finite support. Of course, the belief of a finite support offspring could be accommodated setting $G_0$ accordingly. However, the resulting $DP$ prior would coincide with the usual Dirichlet prior (see Table~\ref{tab1}).


\subsection{Results for complete data}
We first consider a GW process with finite offspring support $\mathcal{S}=\{0,1,\cdots, k\}$. In particular, we set $k=3$ and $\pi=(0.4,0.3,0.2,0.1)$, which yields a critical process ($m=1$). Small variations of these probability values lead to subcritical and supercritical cases, allowing us to test the sensitivity of our estimation methods at detecting the critical threshold. The scenarios and corresponding results are reported in Table~\ref{tab:2}. 

The mle, the DP prior with a low concentration parameter $a$, and the Dirichlet prior exhibit comparable classification performance in most scenarios. However, as the offspring mean approaches the critical value of $m = 1$ in the supercritical case, the performance of all methods deteriorates significantly. In fact, reliable classification is only achieved when $m \geq 3$, highlighting the challenge of accurately identifying processes that are only slightly supercritical. Interestingly, when $a$ is small, the DP prior adapts well to the data, , effectively recovering the true offspring distribution even when the base measure $G_0$ is misspecified as a Poisson distribution. In terms of estimation variability, the mle exhibits the highest variability among the methods considered. In contrast, the DP prior with low $a$ and the Dirichlet prior display similar and more moderate levels of variability, suggesting that these Bayesian approaches provide more stable inference compared to the mle.

The DP prior with large $a$ exhibits behavior similar to the improper prior. Both perform well in the (sub)critical case but struggle significantly in the supercritical case. When the value of $a$ is large relative to the sample size, the DP posterior is dominated by the base measure $G_0$, effectively ignoring the observed data. Consequently, the posterior mean of $m$ is always estimated close to the mean of $G_0$, which is less than one due to the agnostic characterization of the prior (see Table~\ref{tab1}). This suggests that the improper prior implicitly places substantial weight on (sub)critical processes. Indeed, the Chi-square approximation for computing $P(m>1)$ is not reliable in small samples. Consequently, in small samples, a more effective alternative is to use the posterior mean, which coincides with


\begin{table}[ht]     
%
%
\resizebox{\textwidth}{!}{
\begin{tabular}{p{3.5cm}p{3cm}p{3cm}p{2.5cm}p{2.5cm}}
\hline
 & subcritical  & critical &  supercritical&   \\
 Estimator & $m=0.9$  & $m=1$  &  $m=1.2$ & $m=1.5$    \\
\hline
mle               & 0.924 (0.165)  &  0.840 (0.195)   &  0.518 (0.163) & 0.790 (0.235) \\
bayes.improper           & 1.000 (0.04)  & 1.000 (0.071)  & 0.000 (0.112) &  0.000 (0.128)\\
bayes.dir   & 0.924 (0.013)  &  0.840 (0.195)   & 0.594 (0.134)  & 0.794 (0.086) \\
bayes.dp ($a=1$)     & 0.942 (0.063)  & 0.858 (0.084)  & 0.518 (0.099) & 0.790 (0.098)  \\
bayes.dp ($a=100$)     & 1.000 (0.0009)  & 1.000 (0.003)  & 0.020 (0.009)  & 0.008 (0.004) \\
\hline
\end{tabular}
}
\caption{Simulations with complete data. Performance of Finite-support GW process with known $k$. 
Proportion of correct classifications across 500 Monte Carlo replications. 
Parentheses indicate the standard error of the offspring average estimate. 
Labels: \textit{mle} denotes the maximum likelihood estimate; 
\textit{bayes.improper} refers to the Bayesian estimate the using Heyde's improper prior; 
\textit{bayes.dir} represents the Bayesian estimate with a non-informative Dirichlet prior; 
and \textit{bayes.dp} corresponds to the Bayesian estimate with the agnostic $DP(a, G_0)$ prior.}

\label{tab:2}  

\end{table}

Next, we analyze the same dataset as above, but now assuming that $k$ is unknown (Table \ref{tab:3}). Notably, the results for the maximum likelihood and improper Bayesian estimators remain unchanged, as these methods do not require information about $k$. Likewise, the performance of the DP prior is unchanged, since it relies on a Poisson distribution that does not incorporate information about $k$

Mendoza and Guti\'errez-Pe\~na ~\cite{gutpena} do not contemplate the use of the Dirichlet prior when $k$ is unknown. However, a simple solution for using the Dirichlet prior when $k$ is unknown is to estimate $k$ with the sample maximum. The properties of the resulting estimator are unknown and the simulation can provide insight into its performance. As shown in Table \ref{tab:3}, the classification accuracy of the Dirichlet prior is slightly worse than in the case where $k$ is known, reflecting the loss of information.

Thus, when $k$ is unknown, the DP prior exhibits superior classification performance compared to the Dirichlet prior. Additionally, the DP prior allows us to estimate the unknown support size.  This is achieved by estimating $|\mathcal{S}|$ as the modal number of distinct points in 100 posterior samples. The last row of Table \ref{tab:3} reports the proportion of correct estimates, i.e., the fraction of samples where $\widehat{|\mathcal{S}|} = 4$. This proportion increases as more data become available, improving the estimation in supercritical processes. For (sub)critical processes, we observe a trade-off between process classification and support size estimation. With fewer observations, the distribution of $K_N$ tends to be concentrated on smaller values, leading to an underestimation of the support size. At the same time, a low sample size makes it more likely  that the process is correctly classified as (sub)critical.

\begin{table}[ht]
%
%
\resizebox{\textwidth}{!}{
\begin{tabular}{p{4.5cm}p{2.5cm}p{2.5cm}p{2.5cm}p{2.5cm}}
\hline
 & subcritical  & critical &  supercritical&   \\
      & $m=0.9$ & $m=1$ &  $m=1.2$ & $m=1.5$    \\
  Estimator &  &  &   &     \\
\hline
mle               & 0.924 (0.165)  &  0.840 (0.195)  &  0.518 (0.163) & 0.790 (0.235) \\
bayes.improper           & 1.000 (0.04)   & 1.000 (0.071)  & 0.000 (0.112) & 0.000 (0.125)\\
bayes.dir    & 0.918 (0.171)  &  0.828 (0.203)  &  0.532 (0.166) &  0.778 (0.093) \\
bayes.dp ($a=1$)     & 0.942 (0.061)  & 0.858 (0.081)  & 0.518 (0.097)  &  0.786 (0.092)\\
bayes.dp ($a=100$)     & 1.000 (0.001)  & 1.000 (0.003)  & 0.022 (0.008)  &  0.009 (0.003) \\
$\widehat{|\mathcal{S}|} \quad (a=1) $          & 0.170  & 0.242  &  0.442 & 0.778\\
 $\widehat{|\mathcal{S}|} \quad (a=100) $          & 0.924  & 0.932  &  0.926 & 0.970\\
\hline
\end{tabular}
}
\caption{Simulations with Complete Data.  
Performance evaluation based on simulations of data generated from a finite-support GW process, with estimation performed under the assumption that $k$ is unknown.  
Proportion of correct classifications across 500 Monte Carlo replications.  
Parentheses indicate the standard error of the offspring mean estimate.}
\label{tab:3} 
\end{table}

Finally, we consider the GW process with Poisson-distributed offspring. The results are presented in Table \ref{tab:4}, where standard errors are omitted, as they are essentially similar to previous cases. As in prior analyses, the Dirichlet prior requires an estimate of $k$, which we obtain using the sample maximum. The performance of the Dirichlet prior is slightly worse than mle and the DP prior, in particular for supercritical processes. While the mle and DP prior yield comparable results overall, the DP prior demonstrates superior performance in supercritical processes with large values of $m$.
\begin{table}[ht]
%
%
\resizebox{\textwidth}{!}{
\begin{tabular}{p{4cm}p{4cm}p{2cm}p{2cm}p{2cm}}
\hline
 & subcritical  & critical &  supercritical&   \\
      & $m=0.9$ & $m=1$ &  $m=1.2$ & $m=1.5$   \\
  Estimator &  &  &   &     \\
\hline
mle               & 0.946  &  0.858   & 0.344  & 0.514 \\
bayes.improper           & 1.000   &  1.000 & 0.000 &  0.000\\
bayes.dir    & 0.936  &  0.830  &  0.346 &  0.514 \\
bayes.dp ($a=1)$     & 0.940  & 0.866  & 0.322  & 0.620 \\
bayes.dp ($a=100$)    & 1.000  &  1.000 &  0.048 & 0.410 \\
\hline
\end{tabular}
}
\caption{Simulations with Complete Data. Performance evaluation based on simulations of data generated from a GW process with Poisson-distributed offspring. Proportion of correct classifications over 500 MC replications.}
\label{tab:4}  
\end{table}

%
%


\subsection{Results with incomplete data}
\label{res:inco} 

In this section, we present results obtained using only the incomplete data vector. This constraint affects both the Dirichlet and DP prior analyses, as estimating $m$ requires the Gibbs sampler described in Section \ref{algoincomplete}.  

The results for cases where the support size is known are shown in Table \ref{tab:5}, while those where the support size is unknown appear in Table \ref{tab:6}. In both cases, the Dirichlet prior performs similarly to the DP prior. However, the DP prior achieves slightly better results for subcritical processes, whereas the Dirichlet prior exhibits better performance in supercritical processes.  

Comparing Tables \ref{tab:2} and \ref{tab:3} with Tables \ref{tab:5} and \ref{tab:6}, we observe that the DP and Dirichlet prior estimators exhibit reduced performance relative to cases with complete data.  

Finally, we examine incomplete data generated from a Poisson GW process (Table \ref{tab:7}). In this setting, the DP prior outperforms the Dirichlet prior.

\begin{table}[ht]
%
%
\resizebox{\textwidth}{!}{
\begin{tabular}{p{4cm}p{4cm}p{2cm}p{2cm}p{2cm}}
\hline
& subcritical  & critical &  supercritical&   \\
 Estimator &  &  &  $m=1.2$ & $m=1.5$    \\
\hline
mle    & 0.924  &  0.840   & 0.518 & 0.790 \\
bayes.improper &  1.000  & 1.000  & 0.000 & 0.000 \\
bayes.dir (mean) & 0.916  & 0.824  & 0.572 &  0.806\\
bayes.dp ($a=1$) & 0.942 & 0.848  & 0.508  &  0.785\\
\hline
\end{tabular}
}
\caption{Simulations with incomplete data.  
Finite-support GW process with incomplete data, assuming known $k$.  
Proportion of correct classifications across 500 Monte Carlo replications.}
\label{tab:5} 
\end{table}

\begin{table}[ht]
%
%
\resizebox{\textwidth}{!}{
\begin{tabular}{p{4cm}p{4cm}p{2cm}p{2cm}p{2cm}}
\hline
& subcritical  & critical &  supercritical&   \\
 Estimator &  &  &  $m=1.2$ & $m=1.5$    \\
\hline
mle    & 0.924  &  0.840   & 0.518 & 0.790 \\
bayes.improper &  1.000  &  1.000 &  0.000 &  0.000\\
bayes.dir (mean) & 0.922  &  0.84 & 0.678 &  0.822\\
bayes.dp ($a=1$) & 0.932 &  0.854 & 0.604  & 0.808 \\
\hline
\end{tabular}
}
\caption{Simulations with incomplete data.  
Finite-support GW process with incomplete data, assuming unknown $k$ (set to $k=10$).  
Proportion of correct classifications across 500 Monte Carlo replications.}
\label{tab:6} 
\end{table}

\begin{table}[ht]
%
%
\resizebox{\textwidth}{!}{
\begin{tabular}{p{4cm}p{4cm}p{2cm}p{2cm}p{2cm}}
\hline
& subcritical  & critical &  supercritical&   \\
 Estimator &  &  &  $m=1.2$ & $m=1.5$    \\
\hline
mle               & 0.946  &  0.858   & 0.344  & 0.514 \\
bayes.improper &  1.000 & 1.000 & 0.000  & 0.000 \\
bayes.dir (mean) &  0.904 & 0.876 & 0.338 &  0.507\\
bayes.dp ($a=1$) & 0.916 & 0.888 & 0.326  & 0.628 \\
\hline
\end{tabular}
}
\caption{Simulations from a Poisson GW Process with Incomplete Data.  
Finite-support Poisson GW process with incomplete data, assuming unknown $k$ (set to $k=10$).  
Proportion of correct classifications across 500 Monte Carlo replications.}
\label{tab:7}    
\end{table}

\section{Case-study: Sardinia COVID-19 data}
\label{sec:5}
\noindent 
One of the main applications of the GW model is analyzing the spread of epidemics within a population of susceptible individuals. Here, we present a case study on the classification of COVID-19 infectious outbreaks in Sardinia, Italy. The objective is to determine whether the epidemic will eventually extinguish without infecting the entire population. Additionally, we estimate the extinction probability for processes classified as supercritical.  

Several variations of the basic GW model exist for studying epidemic spread. Foster et al. \cite{foster} account for the progressive reduction of the susceptible population in a finite population, which clearly contradicts the assumption of a fixed offspring in the basic GW process. Other authors extended the model to allow for migration and emigration, see for example Kaplan \cite{kaplan}. In the following, we assume that the basic GW model provides a reasonable model for describing epidemic spread in the dataset considered here. 
This assumption appears plausible given the distinctive features of the data. The population size is large, and the data cover a period in which circulation was severely restricted by national legislation due to previous severe outbreaks in other regions, effectively minimizing movement to and from the insular region. Furthermore, the infection waves examined affected only a small portion of the population, making the assumption of a fixed offspring reasonable.  \\

\begin{figure}[h]
\centering
\includegraphics[width=0.95\textwidth]{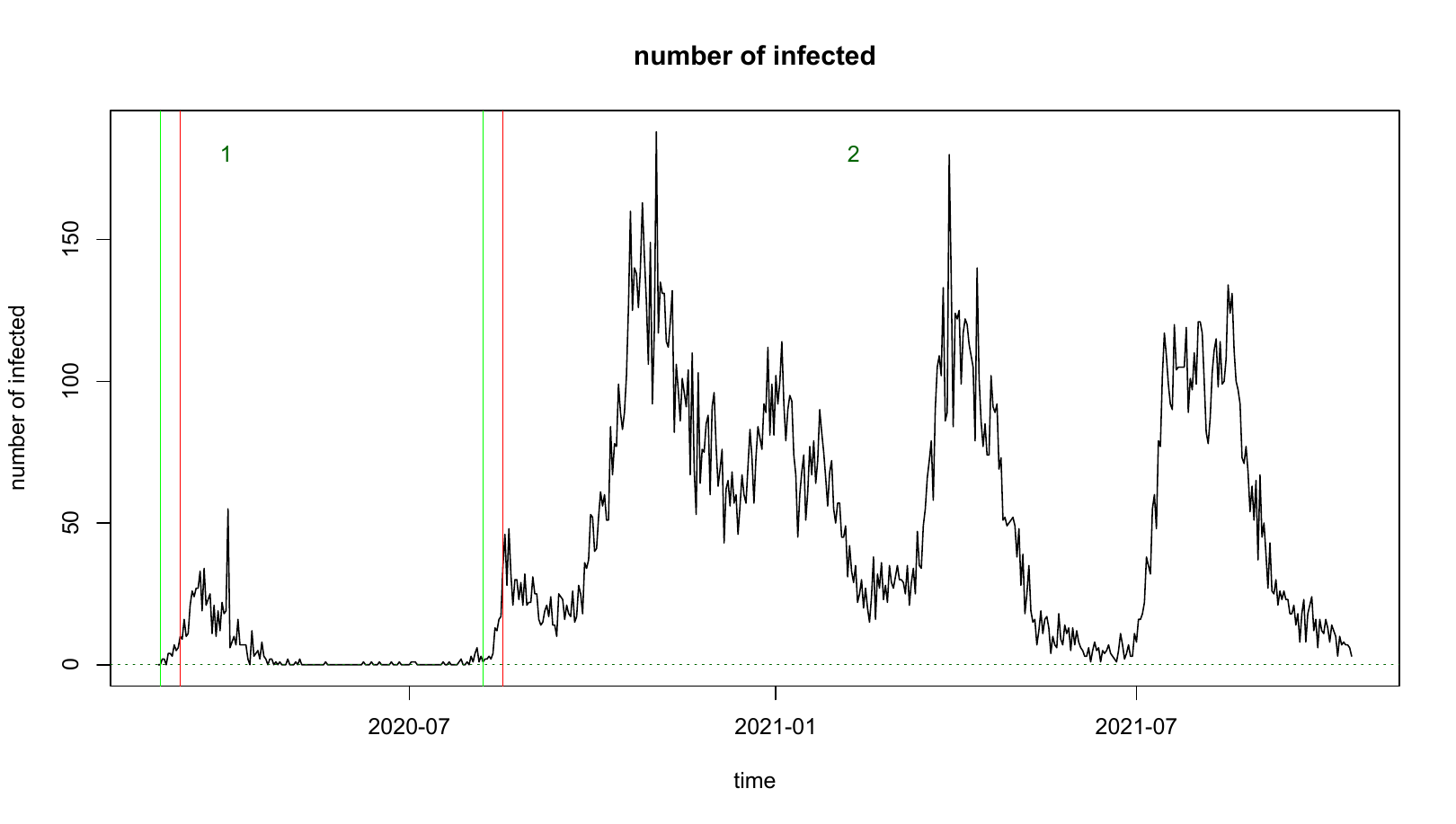}
\caption{Spread of COVID-19 in Sardinia. The numbers indicate the two main waves. The green line indicates the observed starting point of each wave, while the red line denotes the last observed day of each wave. The red and green lines are positioned $10$ days apart.}
\label{figwaves}       
\end{figure}

{\bf Data.}
The data consist of the total number of infected individuals recorded daily over a period of approximately two years, from February 26, 2020, to February 20, 2022, in the Italian region of Sardinia. Figure~\ref{figwaves} displays the number of infected individuals (excluding hospital-working physicians) as a function of time.   
 
 Several infection waves are clearly identifiable. The first wave is  clearly separated from the next, and smaller than the subsequent waves. The second wave, which lasted several months,  is multimodal and affected a larger number of individuals.

 As is common in epidemiological studies, the dataset is incomplete: while the total number of infected individuals is observed during each period, the number of individuals infected by each person is unknown. To address this limitation, we implemented the algorithm described in Section \ref{res:inco}, enabling the application of both the DP and Dirichlet priors to incomplete data. Thus, we implemented the algorithm described in section \ref{res:inco} when using both the DP and the Dirichlet prior with incomplete data.\\

\subsection{Results}
In this section, we present estimation results using maximum likelihood and Bayesian analysis with Dirichlet, improper, and DP priors on the offspring distribution. The $DP(a, G_0)$ prior adopts an agnostic base measure, $G_0 \sim \text{Poisson}$, with $a = 1$ for the offspring distribution.  

Of particular importance is the model's early detection ability, that is, its ability to accurately predict the eventual extinction using data from the very initial periods. To evaluate this, we present estimates derived from the first $n$ days of the outbreak ($n = 2, 4, 6, 8, 10$). 

Using early--stage data mitigates the issue of progressive reduction in the susceptible population. Moreover, the regulatory context during the waves is one of severe restrictions due to previous large outbreaks in other regions, making the assumptions of the basic GW model, e.g., no immigration or emigration and stable offspring, more plausible.

For the first wave (Table~\ref{tab:8}) the estimates of the offspring average $m$ suggest that the process is supercritical. Additionally, all estimates are slightly greater than $1$, a value much smaller than the theoretical value of $m$ for COVID-19 spread in absence of restrictions, which is generally considered to be approximately $2.5$, see Dhunghel et al. \cite{covid19}. 
These estimates reflect the stringent measures in place during the first wave, which were aimed at slowing the infection’s spread.  Since $m$ is only slightly greater than $1$, the extinction probability is intuitively high,  making wave extinction a reasonable expectation.  

The estimates of the extinction probabilities assuming geometric offspring are shown in Table \ref{tab:8prob}. Estimates assuming a Poisson offspring distribution yield very similar results (not shown).  

\begin{table}[ht]
%
%
\resizebox{\textwidth}{!}{
\begin{tabular}{p{6cm}p{2cm}p{2cm}p{2cm}p{2cm}p{1cm}}
\hline
data up to day $n$ & 2  & 4 &  6 & 8 & 10 \\
                  &  &  &   &  &    \\
\hline
mle               & 2.000  & 1.000   & 1.300  & 1.353 & 1.172 \\
bayes.improper ($P(m>1)$) & 0.736  & 0.440  & 0.792  & 0.905 & 0.801 \\
bayes.dir (mean) & 2.052  & 1.019 & 1.297 & 1.349 & 1.171 \\
bayes.dp ($G_0$ Poisson) & 1.384  & 0.994  & 1.261  & 1.325 & 1.157\\
\hline
\end{tabular}
}
\caption{Spread of COVID-19 in Sardinia. 
Estimates of the offspring averaged based on the observed infection dynamics during the first wave.}
\label{tab:8}  
\end{table}

\begin{table}[ht]
%
%
\resizebox{\textwidth}{!}{
\begin{tabular}{p{6cm}p{2cm}p{2cm}p{2cm}p{2cm}p{1cm}}
\hline
data up to day $n$ & 2  & 4 &  6 & 8  & 10 \\
                  &  &  &   &  &    \\
\hline
mle               & 0.500  & 1.000   & 0.769  & 0.739 & 0.853 \\
bayes.dir (mean) & 0.487  & 0.981 & 0.771 & 0.741 & 0.854 \\
bayes.dp ($G_0$ Poisson) & 0.723  & 1.000  & 0.793  & 0.755 & 0.864\\
\hline
\end{tabular}
}
\caption{Spread of COVID-19 in Sardinia. Extinction probabilities  based on the observed infection dynamics during the first wave of the COVID-19 epidemic in Sardinia considering a geometric offspring.}
\label{tab:8prob}  
\end{table}

The estimates of the offspring average for the second wave are presented in Table~\ref{tab:9} and are generally higher than those for the first wave, reflecting the larger scale of the second outbreak. Consequently, the extinction probabilities are lower (see Table~\ref{tab:9prob}). In both waves, the DP prior estimates are slightly lower than those obtained from the Dirichlet prior. This difference arises because the DP prior assigns a small but positive probability to the tail of the Poisson offspring distribution.  

\begin{table}[ht]
%
%
\resizebox{\textwidth}{!}{
\begin{tabular}{p{6cm}p{2cm}p{2cm}p{2cm}p{2cm}p{1cm}}
\hline
data up to day $n$ & 2  & 4 &  6 & 8  & 10 \\
                  &  &  &   &  &    \\
\hline
mle               &  2.000 &  1.400  & 1.300  & 1.407  &  1.291\\
bayes.improper ($P(m>1)$) & 0.736 & 0.762  & 0.792  & 0.974 &  0.978\\
bayes.dir (mean) & 2.064  & 1.386 & 1.306 & 1.412 & 1.293 \\
bayes.dp ($G_0$ Poisson) & 1.504 & 1.357  & 1.242 & 1.371 & 1.280\\
\hline
\end{tabular}
}
\caption{Spread of COVID-19 in Sardinia. Estimates of the offspring average based on the second wave of the COVID-19 epidemic in Sardinia.}
\label{tab:9} 
\end{table}

\begin{table}[ht]
%
%
\resizebox{\textwidth}{!}{
\begin{tabular}{p{6cm}p{2cm}p{2cm}p{2cm}p{2cm}p{1cm}}
\hline
data up to day $n$ & 2  & 4 &  6 & 8  & 10 \\
                  &  &  &   &  &    \\
\hline
mle               & 0.500  & 0.714  & 0.769  & 0.711 & 0.775 \\
bayes.dir (mean) & 0.484  & 0.722 & 0.766 & 0.708 & 0.773 \\
bayes.dp ($G_0$ Poisson) & 0.665  & 0.737  & 0.805  & 0.729 & 0.781\\
\hline
\end{tabular}
}
\caption{Spread of COVID-19 in Sardinia. Extinction probabilities based on the second wave of the COVID-19 epidemic in Sardinia considering a geometric offspring.}
\label{tab:9prob}  
\end{table}

%
%


\section{Conclusions}
\label{sec:6}
\noindent
 In this paper, we described a Bayesian nonparametric approach for inference on the GW model, proposing a Dirichlet Process prior for the offspring distribution. An advantage of the $DP(a, G_0)$ prior is its flexibility, allowing prior beliefs about the offspring distribution to be set in a very natural way by choosing $G_0$ as the prior guess and $a$ to reflect confidence in this assumption. Moreover, the DP prior does not require knowledge of the offspring support size, which can be estimated directly from the data. We also showed that the DP prior can be made agnostic (noninformative) with respect to the key parameter $m$, the offspring average.  
 
 We investigated the small-sample performance of the DP prior with the aid of a simulation study. These simulations allowed us to compare the DP prior with maximum likelihood estimation (MLE) and Bayesian estimators. While competing methods exhibit strong asymptotic properties, previous studies had not examined their small-sample behavior. Our results indicate that the DP prior, with a low concentration parameter and either a Poisson or Geometric base measure, performs well in process classification. However, the support size tends to be underestimated unless the sample size is sufficiently large, particularly in supercritical processes. The  performance of the DP prior deteriorates only slightly when complete data are not available.

 Bayesian nonparametric inference is generally computationally intensive. Some posterior summaries, such as mean estimates, can be obtained efficiently, but inference on the offspring support size requires posterior sampling. Estimation under incomplete data also necessitates a Gibbs sampling approximation. To address these challenges, we employed computationally efficient methods: the Doss algorithm \cite{doss} for DP posterior sampling and a Blocked Gibbs sampler for the DP prior with incomplete data. While the computational burden is substantial for large datasets, for GW generated data, this burden is significantly reduced because inference is typically required only for data representing the few first generations of an outbreak. Additionally, the complexity is reduced for (sub)critical processes, which tend to extinguish after a few generations.  

 Several promising avenues exist for future research. This paper focused on the simplest GW processes with Poisson, geometric, and finite-support offspring distributions. A more realistic approach may assume population heterogeneity by partitioning individuals into subgroups with distinct offspring distributions, leading to a mixture model whereby the data generated by this offspring are over dispersed compared to a common offspring for all individuals.  A natural nonparametric extension involves placing a DP prior on the unknown mixing distribution. We leave this promising extension for future investigation. 
\paragraph{Acknowledgments}
Research partially supported by the project UniCA/FdS 2020 by Fondazione di Sardegna

\paragraph{Data Availability} The data used in the application is available at the link: \url{https://covid19.infn.it/iss/}

\section*{Declaration}
\paragraph{Ethical Approval}
This research does not contain any studies with human participations or animals performed by any of the authors.
\paragraph{Conflict of Interest}
The authors declare no competing interests.

\section*{Appendix}
\addcontentsline{toc}{section}{Appendix}

\begin{figure}[h]
\centering
\includegraphics[width=0.9\textwidth]{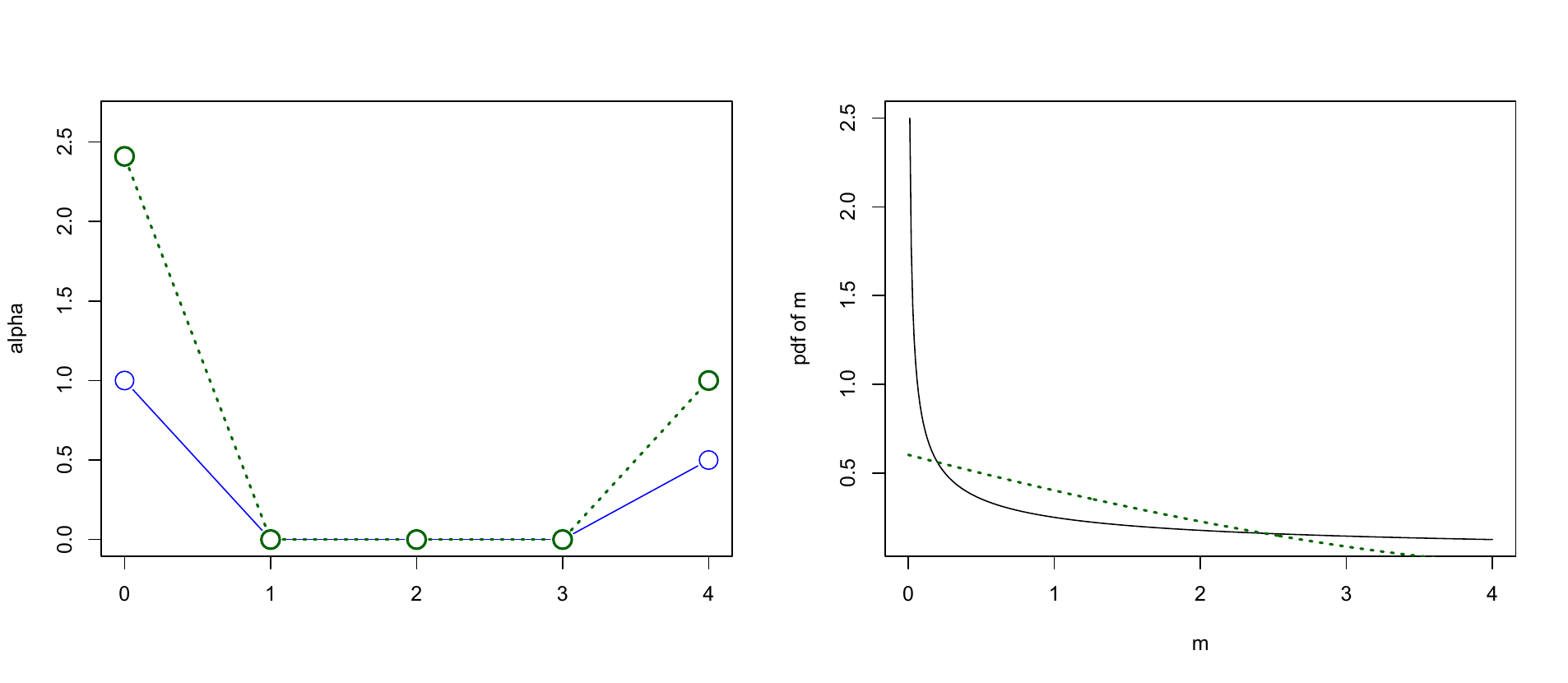}
\caption{Agnostic prior for $\pi$ (left) and corresponding prior induced on $m$ (right). Continuous line is the Mendoza and Guti\'errez-Pe\~na prior; dotted line is another agnostic prior.}
\label{fig:neutral}       
\end{figure}

 In this Appendix we recover the uninformative prior for $m$ proposed by Mendoza and Guti\'errez-Pe\~na ~\cite{gutpena}. We also show that this prior is not unique and derive another agnostic prior (see Figure~\ref{fig:neutral}). The rationale behind the agnostic prior is extended in section~\ref{sec:3} to obtain an agnostic DP prior.

First of all, we observe that any flat Dirichlet prior on $\pi= (\pi_0, \cdots, \pi_k)$ induces a prior mean on $m$ which increases with $k$ so it is not uninformative on $m$. An uninformative prior on $m$ should assign the same prior probability to sub and supercritical processes so we must have $Pr(m>1)=1/2$. Mendoza and Guti\'errez-Pe\~na (\cite{gutpena}, section 3.2) showed that an approximate non-informative prior for $m$ is a Dirichlet prior with parameters $\alpha_0=1,\alpha_1\approx 0, \cdots,\alpha_{k-1}\approx 0,\alpha_k=\frac{log(2)}{log(k)}$. These values are derived as follows. Let $\theta_j=log(\frac{\pi_j}{\pi_0+\pi_k})$ so that $\theta_j$ follows a generalized logistic distribution, $j =1, \cdots, k-1$. The generalized logistic is unimodal, with a priori mode given by $\log(\alpha_j)-\log(\alpha-\sum_{l=1}^{k-1}\alpha_l)$. It turns out that this transformation induces on $p(m)=p(m(\theta))$ a non-standard beta distribution: 

$$p(m)=\frac{(m-c)^{a-1}(d-m)^{b-1}}{B(a,b)(d-c)^{a+b-1}}$$

where $a=\alpha_k$, $b=\alpha_0$, $c=\sum_1^{k-1}j\frac{\alpha_j}{\sum \alpha}$, $d=k-\sum_1^{k-1}(k-j)\frac{\alpha_j}{\sum \alpha}$. \\

For this prior to be uninformative we impose that its median is equal to the critical value of one. If we set $\alpha_0=1$, $\alpha_1\approx 0, \cdots, \alpha_{k-1}\approx 0$ the median of the distribution depends only on $\alpha_k$:
$$ median( m) = d\cdot median(Beta(a,b)) = d(1/2)^{1/a}\approx k(1/2)^{1/a}$$ 
and solving the equation $median(m)=1$ we find $\alpha_k=\frac{log(2)}{log(k)}$.

Finally, we note that another uninformative prior can be obtained setting $\alpha_k=1$ instead of $\alpha_0$. If so, we have $median(m)=1-(1/2)^{1/b}$ and proceeding analogously we find $\alpha_0=\frac{log(2)}{log(\frac{k}{k-1})}$. Both priors are shown in Figure~\ref{fig:neutral} for the case $k=4$. 

\end{document}